\documentclass[prb,aps,twocolumn,amstex,preprintnumbers,amsmath,amssymb,array,tabularx]{revtex4}

\usepackage{graphicx}
\usepackage{dcolumn}
\usepackage{bm}

\DeclareGraphicsExtensions{.jpg,.pdf,.eps}

\begin{document}
\title{Mode-coupling theory for reaction dynamics in liquids}
 
\author{Nurit Shental and Eran Rabani\footnote{Author to whom
correspondence should be addressed; electronic mail:
rabani@tau.ac.il}}
\address{School of Chemistry, The Sackler Faculty
of Exact Sciences, Tel Aviv University, Tel Aviv 69978, Israel}

\date{\today}

\begin{abstract}
A theory for chemical reaction dynamics in condensed phase systems
based on the generalized Langevin formalism of Grote and Hynes is
presented.  A microscopic approach to calculate the dynamic friction
is developed within the framework of a combination of kinetic and
mode-coupling theories.  The approach provides a powerful analytic
tool to study chemical reactions in realistic condensed phase
environments.  The accuracy of the approach is tested for a model
isomerization reaction in a Lennard-Jones fluid.  Good agreement is
obtained for the transmission coefficient at different solvent
densities, in comparison with numerical simulations based on the
reactive-flux approach.
\end{abstract}

\maketitle
 
\newpage 

\section{Introduction}
\label{sec:intro}
Chemical reactions in condensed phases are frequently described using
two different methods.\cite{Hanggi90} The simple approach consists of
a Brownian particle moving in a one-dimensional bistable potential.
In this case the dynamics of this system can be described by the
Langevin equation.\cite{Kramers40} Grote and Hynes~\cite{Grote80} have
extended this approach to non-Markovian processes for a parabolic
barrier, where the dynamics are given by the {\em generalized}
Langevin equation (GLE).\cite{Zwanzig61} These and other theories were
tested numerically by Straub, Borkovec and
Berne,\cite{Berne85a,Berne86} and their results were crucial in the
development of the Pollak, Grabert, H{\"{a}}nggi turnover
theory.\cite{Pollak89}

The other approach, which became more common, is based on molecular
dynamics simulations, and can account for a general form of the
Hamiltonian.  Special methods have been developed to accelerate the
barrier crossing and thereby numerically determine the rate constant.
Chandler~\cite{Chandler78} showed that in the time-correlation
approach to rate constants,\cite{Anderson73,Bennett77} the reactive
flux rapidly decays to a plateau value, which can then be associated
with the slow rate for crossing the barrier.\cite{Montgomery79} For
cases in which no preconceived notion of mechanism or transition state
is known, a more general computational method, called the ``transition
path sampling,'' has recently been developed and applied to complex
reactive systems.\cite{Chandler98,Chandler01,Chandler02,Dellago02}

One of the major limitations of the simple approach based on the GLE
formalism is related to how one can evaluate the frequency dependent
memory friction (related to the random force by the fluctuation
dissipation theorem).  Most applications assume a Gaussian random
force for the GLE. In this limit, the dynamics can be transformed into
a Hamiltonian description where the system is {\em linearly} coupled
to a harmonic bath.\cite{Zwanzig73,Lindenberg85} Bagchi and his
coworkers have used a mode-coupling
theory~\cite{BalucaniZoppi,BoonYip,HansenMcDonald} to generalize this
approach in order to include non-Gaussian
fluctuations.\cite{Bagchi97,Bagchi99} They assumed that the
translational friction is the principal quantity that regulates the
diffusive Brownian motion of the reactive system near the barrier
region.  In this limit, the memory friction is simply given by the
Sj{\"{o}}gren and Sj\"{o}lander mode-coupling expression for the
memory kernel of a GLE for self-motion in neat
liquids.\cite{Sjogren79}

In this paper we develop an alternative theory to treat non-Gaussian
fluctuations within the GLE formalism.  First, following the work of
Oppenheim and his coworkers,\cite{Oppenheim70,Oppenheim82,Oppenheim91}
we derive a GLE for the dynamics of a reactive system in a liquid
host.  Based on the formal expression for the memory friction of this
GLE, and using a combination of kinetic and mode-coupling theories, we
obtain a simple expression for the memory friction.  This memory
friction is then used to obtain the reaction rate within the framework
of the Grote-Hynes theory.  Our approach is different from that of
Bagchi and coworkers~\cite{Bagchi97,Bagchi99} in two ways.  First, we
need not assume that the translational friction is the principal
quantity that regulates the diffusive Brownian motion of the reactive
system.  Second, it can be shown that this assumption limits the form
of the couplings between the system and the liquid host.  Thus, a more
general form of the Hamiltonian can be treated within our formulation.

Our paper is organized as follows: In Section~\ref{sec:theory} we
provide a derivation of the GLE, an overview of the Grote-Hynes
theory, and a derivation of our kinetic and mode-coupling theories.
Tests of our approach for a model isomerization reaction in a LJ fluid
is provided in Section~\ref{sec:results}.  Finally, we conclude in
Section~\ref{sec:conclusions}.

\section{Theory}
\label{sec:theory}
The development of our approach to reaction dynamics is based on three
steps.  First, following the work of Oppenheim and his
coworkers,\cite{Oppenheim70,Oppenheim82,Oppenheim91} we derive a GLE
for the dynamics of a subsystem.  Second, we adopt the Grote-Hynes
theory,\cite{Grote80} which relates the reaction rate to the memory
kernel of the GLE.  Finally, we develop a theory based on a
combination of kinetic and mode-coupling
approaches~\cite{BalucaniZoppi,BoonYip,HansenMcDonald} to calculate
the reaction rate within the Grote-Hynes formalism.

\subsection{Generalized Langevin Equation}
\label{subsec:gle}
Consider a general Hamiltonian for a reactive system in a liquid host
of the from
\begin{equation}
H = H_{s}({R},{P}) + H_{b}({\bf r},{\bf p}) + \phi({R},{\bf r}),
\label{eq:hamil}
\end{equation}
where $H_{s}({R},{P})$ is the Hamiltonian of a reactive one
dimensional system with phase-space coordinates ${R}$ and ${P}$, and
reduced mass $\mu$, $H_{b}({\bf r},{\bf p})$ is the Hamiltonian of $N$
interacting liquid particles with phase-space coordinates ${\bf r}$
and ${\bf p}$, and mass $m$, and $\phi({R},{\bf r})$ is the coupling
between the system and the solvent.

Using the projection operator formalism of Zwanzig and
Mori,\cite{Zwanzig61,Mori65a,Mori65b} and following the work of Mazur
and Oppenheim,\cite{Oppenheim70} we now derive reduced equations of
motion for the system.  First, we define a projection operator ${\cal
P}$ that projects out the bath variables:
\begin{equation}
{\cal P} B = \int d{\bf r} d{\bf p}\bar{\rho}({R},{\bf r},{\bf p})
B \equiv \langle B \rangle,
\label{eq:projection}
\end{equation}
where
\begin{equation}
\bar{\rho}({R},{\bf r},{\bf p}) = \rho_{b}({\bf r},{\bf p})
\exp\{-\beta [\phi({R},{\bf r})-w({R})] \}.
\label{eq:rhohat}
\end{equation}
In the above equation $\beta=1/k_{B}T$ is the inverse temperature,
$\rho_{b}({\bf r},{\bf p})=\exp\{-\beta H_{b}({\bf r},{\bf
p})\}/Z_{b}$ is the equilibrium distribution function for the isolated
bath, $Z_{b}$ is the partition function of the isolated bath, and
$w({R})$ is the potential of mean force, given by:
\begin{equation}
\exp\{-\beta w({R})\} = \int d{\bf r} d{\bf p} \rho_{b}({\bf r},{\bf
p}) \exp\{-\beta \phi({R},{\bf r}) \}.
\label{eq:wr}
\end{equation}

Using the above projection operator it is straightforward to show that
the reduced equations of motion for the system are given by the GLE
from:\cite{Oppenheim70}
\begin{equation}
\dot{R}=\frac{P}{\mu},
\label{eq:rdot}
\end{equation}
and
\begin{equation}
\begin{split}
\dot{P}=-\mbox{e}^{i L t} &\left[\frac{\partial H_{s}}{\partial {R}} -
\frac{\partial w}{\partial {R}}\right] + K(t) \\ &- \frac{\beta}{\mu}
\int_{0}^{t} d\tau \mbox{e}^{i L \tau} P \langle K K(t-\tau) \rangle,
\end{split}
\label{eq:pdot}
\end{equation}
where, as noted above, $\mu$ is the reduced mass of the system, the
$\langle \cdots \rangle$ is defined in Eq.~(\ref{eq:projection}),
$\mbox{e}^{i L t}$ is the classical propagator with $iL=\{H,\cdots\}$,
and the random force $K(t)$ is given by
\begin{equation}
K(t)=-\mbox{e}^{i (1-{\cal P}) L t} \frac{\partial}{\partial {R}}
(\phi({R},{\bf r})-w({R})).
\label{eq:randomforce}
\end{equation}
Note that the random force involves {\em projected} propagation where
the classical propagator $\mbox{e}^{i L t}$ is replaced with the
projected propagator $\mbox{e}^{i (1-{\cal P}) L t}$.

The memory kernel of the above GLE is related to the random force by
the fluctuation dissipation theorem, and is given by
\begin{equation}
\zeta(t) = \frac{\beta}{\mu} \langle K K(t) \rangle.
\label{eq:memorykernel}
\end{equation}
This correlation function is, in general, a function of time and a
function of the system coordinate $R$
(cf. Eq.~(\ref{eq:randomforce})).

\subsection{Grote Hynes Theory}
\label{subsec:ght}
To obtain the reaction rate one needs to solve the GLE given by
Eq.~(\ref{eq:pdot}).  However, as a result of the fact that the random
force involves projected propagation of the couplings between the
system and the bath, a complete solution of this GLE is an extremely
difficult task.  To circumvent this problem, Grote and Hynes have
developed an approximate theory to obtain the rate within the GLE
formulation.\cite{Grote80}

The basic assumption made by Grote and Hynes is that in the {\em
barrier region}, the dynamics can also be described by the above GLE,
where the deterministic force (give by $\left[\frac{\partial
H_{s}}{\partial {R}} - \frac{\partial w}{\partial {R}}\right]$) is
replaced with an inverted parabolic approximation.  After some lengthy
algebra, Grote and Hynes obtained the following simple result for the
transmission coefficient $\kappa$:
\begin{equation}
\kappa = \frac{k}{k_{TST}} = \frac{\lambda_{r}}{\omega_{b}}.
\label{eq:ghrate}
\end{equation}
The transmission coefficient is given by the ratio of the reactive
frequency ($\lambda_{r}$) to the barrier frequency ($\omega_{b}$).
The former is related to the memory friction appearing in the GLE, and
is given by
\begin{equation}
\lambda_{r} = \frac{\omega_{b}^{2}}
{\lambda_{r}+\hat{\zeta}^{\dagger}(\lambda_{r})},
\label{eq:lambdar}
\end{equation}
where $\hat{\zeta}^{\dagger}(\lambda_{r})$ is the Laplace transform of
$\zeta^{\dagger}(t)$ given by
\begin{equation}
\hat{\zeta}^{\dagger}(\lambda_{r}) = \int_{0}^{\infty}dt
\mbox{e}^{-\lambda_{r} t} \zeta^{\dagger}(t),
\label{eq:laplace}
\end{equation}
and the symbol ``$\dagger$'' indicates that the observable is
evaluated at the saddle point, namely, in the barrier region as
required by the Grote-Hynes theory.  To obtain the rate
Eq.~(\ref{eq:lambdar}) must be solved self-consistently.

The Grote-Hynes expression for the rate reduces to Kramer's result if
the memory friction decays rapidly on timescales faster than the
system's motion.\cite{Grote80} In the strong friction limit,
$\hat{\zeta}^{\dagger}(\lambda_{r})/\mu >> \lambda_{r}$, the reactive
frequency will be much smaller than the barrier frequency resulting in
a small transmission coefficient.  In the weak friction limit,
$\hat{\zeta}^{\dagger}(\lambda_{r})/\mu << \lambda_{r}$, the reactive
frequency simply equals the barrier frequency
($\lambda_{r}=\omega_{b}$), and the rate is given by the transition
state rate.  This is a shortcoming of the approximation made by Grote
and Hynes which fails to capture the turn over to Kramer's spatial
diffusion regime.  Hence, the Grote-Hynes rate expression is valid
only for high frictions.

The approach developed by Grote and Hynes reduces the complexity of
solving the GLE to that of estimating the memory friction of the GLE.
For such problems, a very powerful approach has been developed and
applied to many interesting dynamical problems in solutions.  In the
following subsection we describe a combination of kinetic and
mode-coupling theories that we have developed to model the memory
friction of the GLE in order to study reaction dynamics in liquids.

\subsection{Kinetic and Mode-Coupling Theories}
\label{subsec:mct}
One major difficulty in evaluating the memory kernel of the GLE given
in Eq.~(\ref{eq:pdot}) is that the correlation of the random force
involves propagation of $\phi({R},{\bf r})-w({R})$ in the projected
subspace spanned by $Q=1-{\cal P}$ (see Eq.~(\ref{eq:randomforce})).
To overcome this difficulty we develop a theory to calculate
$\zeta(t)$ that is based on a combination of kinetic and mode-coupling
theories.\cite{BalucaniZoppi} This combination has been used to study
density and current
fluctuations,\cite{BalucaniZoppi,BoonYip,HansenMcDonald} solvation and
relaxation dynamics,\cite{Bagchi99,Reichman02b,Egorov03} and nonlinear
spectroscopy~\cite{Denny01,Denny02a,Denny02b} in classical liquids.  A
quantum mechanical generalization of this approach has recently been
developed and applied successfully to study the dynamic response in
quantum
liquids.\cite{Rabani02a,Reichman01a,Rabani02b,Reichman02a,Rabani02c,Rabani02d,Rabani04}

The Grote-Hynes approach requires that the memory friction be
approximated in the barrier region.  We therefore replace the full
memory friction of the GLE ($\zeta(t)$) with that approximated at the
saddle point ($\zeta^{\dagger}(t)$), where (as before) the symbol
``$\dagger$'' indicates that the position variable of the system is
taken at the saddle point.  Next, we replace $\zeta^{\dagger}(t)$ with
an approximate form given by~\cite{Gotze92}
\begin{equation}
\zeta^{\dagger}(t) = \zeta^{\dagger}_{B}(t) + \zeta^{\dagger}_{MC}(t),
\label{eq:memory}
\end{equation}
where $\zeta^{\dagger}_{B}(t)$ and $\zeta^{\dagger}_{MC}(t)$ are the
``binary'' and ``mode-coupling'' terms of the memory friction,
respectively.  The fast decaying binary term is determined from a
short-time expansion (to second order in time) of the exact memory
friction, and is given by
\begin{equation}
\zeta^{\dagger}_{B}(t) = \zeta^{\dagger}(0) {~} \exp(-(t/\tau)^2),
\label{eq:zetaB}
\end{equation}
where the lifetime $\tau$ is given by
\begin{equation}
\frac{1}{\tau^{2}} = -\frac{1}{2}
\frac{\ddot{\zeta}^{\dagger}(0)}{\zeta^{\dagger}(0)}.
\label{eq:tau}
\end{equation}
In the above equations $\zeta^{\dagger}(0)$ and
$\ddot{\zeta}^{\dagger}(0)$ are the zero and second time moments of
the memory friction, and are given by
\begin{equation}
\zeta^{\dagger}(0) = \frac{\beta}{\mu} \left \langle \left\{
\frac{\partial}{\partial {R}^{\dagger}} (\phi({R}^{\dagger},{\bf
r})-w({R}^{\dagger})) \right\}^{2} \right \rangle,
\label{eq:zeta0}
\end{equation}
and
\begin{equation}
\ddot{\zeta}^{\dagger}(0) = -\frac{1}{m \mu} \left \langle \left\{
\frac{\partial}{\partial {R}^{\dagger}}\frac{\partial}{\partial {\bf
r}} (\phi({R}^{\dagger},{\bf r})-w({R}^{\dagger})) \right\}^{2} \right
\rangle.
\label{eq:d2zeta0}
\end{equation}
As noted above, $\mu$ is the reduced mass of the system, and $m$ is
the mass of a liquid particle.

The slow decaying mode-coupling portion of the memory kernel,
$\zeta^{\dagger}_{MC}(t)$, must be obtained from a mode-coupling
approach.  The basic idea behind this approach is that the random
force projected correlation function decays at intermediate and long
times predominantly into modes that are associated with
quasi-conserved dynamical variables.  It is reasonable to assume that
the decay of the memory kernel at long times will be governed by those
modes that have the longest relaxation time.  In the present
application the slow decay is basically attributed to couplings
between wavevector-dependent density modes of the form
\begin{equation}
b_{{\bf k},{\bf q}} = c_{\bf k} {~} n_{\bf q},
\label{eq:b}
\end{equation}
where the self-density mode is given by
\begin{equation}
c_{\bf k} = \mbox{e}^{i {\bf k} \cdot {\bf r}_{j}},
\label{eq:c}
\end{equation}
with ${\bf r}_{j}$ being the coordinate of liquid particle $j$, and
the density mode is given by
\begin{equation}
n_{\bf k} = \sum_{j=1}^{N} \mbox{e}^{i {\bf k} \cdot {\bf r}_{j}},
\label{eq:n}
\end{equation}
for $N$ liquid particles.

In practice, the simplest way to extract the dominant slow
contribution of the decay of the memory friction is to introduce
another projection operator, $P_{2}$ given by
\begin{equation}
P_{2} = \sum_{{\bf k}{\bf q}} \frac{b_{{\bf k},{\bf q}}}{N S(q)}
\langle b_{{\bf -k},{\bf -q}}, \cdots \rangle,
\label{eq:P2}
\end{equation}
that projects any variables on the space spanned by the slow modes
$c_{\bf k}$ and $n_{\bf q}$.  Then, following the common
approximations in which the projected dynamics of the random force is
replaced with the dynamics projected onto these slow variables, and
replacing four-point density correlations with a product of two-point
density correlations,\cite{BalucaniZoppi,BoonYip,HansenMcDonald} we
find that $\zeta^{\dagger}_{MC}(t)$ is given by
\begin{equation}
\begin{split}
\zeta^{\dagger}_{MC}(t) &= \frac{\beta}{\mu \rho} \int
\frac{dq^{3}}{(2\pi)^{3}} \left[|V^{\dagger}({\bf
q})|^2/S(q)^{2}\right] \\ & \left[ F_{s}(q,t) F(q,t) - F_{sb}(q,t)
F_{b}(q,t)\right].
\end{split}
\label{eq:zetamc}
\end{equation}
In the above equation $\rho$ is the liquid number density, $S(q)$ is
the structure factor of the neat fluid, $F_{s}(q,t)$ and $F(q,t)$ are
the self-intermediate and intermediate scattering functions of the
neat fluid, respectively, and the vertex is given by the static
average of the product of the coupling force and the slow modes:
\begin{equation}
V^{\dagger}({\bf q})=\left\langle \frac{\partial}{\partial
{R}^{\dagger}} [\phi({R}^{\dagger},{\bf r})-w({R}^{\dagger})] b_{{\bf
q},-{\bf q}} \right\rangle.
\label{eq:vertex}
\end{equation}
As before, the symbol ``$\dagger$'' indicates that the coupling force
is evaluated at the saddle point.

The binary self-intermediate and binary intermediate scattering
functions are given by
\begin{equation}
F_{sb}(q,t) = \exp\left\{-\frac{q^2t^2}{2\beta m} \right\}
\label{eq:fsb}
\end{equation}
and
\begin{equation}
F_{b}(q,t) = S(q) \exp\left\{-\frac{q^2t^2}{2\beta m S(q)} \right\}
\label{eq:fb}
\end{equation}
The subtraction of the product of these terms in Eq.~(\ref{eq:zetamc})
is done to prevent over-counting the total memory kernel at short
times, namely, to ensures that the even time moments of the total
memory kernel are exact to forth order in time.

To obtain the memory friction one requires as input the first two
moments $\zeta^{\dagger}(0)$ and $\ddot{\zeta}^{\dagger}(0)$, the
static structure factor $S(q)$ and the vertex $V^{\dagger}({\bf q})$.
These static averages can be obtained from simulations, or from the
proper integral equation formulation.  In addition, we need the
self-intermediate and intermediate scattering functions.  These
time-dependent correlation functions of the neat fluid can be obtained
from simulations, or alternatively from a similar mode-coupling
approach, where a GLE for the density fluctuations is solved using a
combination of kinetic and mode-coupling
theories.\cite{BalucaniZoppi,BoonYip,HansenMcDonald}

\section{Results}
\label{sec:results}
To assess the accuracy of the proposed theory we have studied the rate
of a model for an isomerization reaction of a diatomic molecule in a
Lennard-Jones (LJ) fluid.  The Hamiltonian of the entire system and
bath can be described by Eq.~(\ref{eq:hamil}).  The isomerizing
diatomic molecule is made of two atoms with equal mass $m^{*}$,
interacting via a symmetric double-well potential.  Without loss of
generality, we place the atoms along the $z$-axis, where the position
vectors are ${\bf R}_{1}=\{0,0,0\}$ and ${\bf R}_{2}=\{0,0,R\}$ for
atom $1$ and $2$, respectively. The reaction coordinate is taken as
the distance $R=|{\bf R}_{1}-{\bf R}_{2}|$ separating the atoms.  We
allow the atoms to move only along the $z$ axis (which is the reaction
coordinate).  The reactive system Hamiltonian can be described by:
\begin{equation}
H_{s}({R},{P}) = \frac{{P}^{2}}{2\mu} + V_{0} \{4[(R -
R^{\dagger})/a]^{2} - 1\}^2,
\label{eq:hs}
\end{equation}
where $\mu=m^{*}/2$ is the reduced mass of the system, $V_{0}$ is
energy barrier separating reactants from products, $R^{\dagger}$ is
the location of the transition state, and $a$ is the distance between
the two minima corresponding to stable reactants and products.  For
the results shown below we take $R^{\dagger} = 2^{1/6}\sigma$,
$V_{0}=5k_{B}T$, $T=2.5 \epsilon$, $a=2\sigma/3$, and a reduced mass
$\mu=1/2$ for the isomerizing diatomic molecule.

The solvent Hamiltonian is given by the LJ form:
\begin{equation}
H_{b}({\bf r},{\bf p}) = \sum_{j=1}^{N}\frac{{\bf p}_{j}^{2}}{2 m} +
\sum_{i>j=1}^{N} 4\epsilon \left[
\left(\frac{\sigma}{r_{ij}}\right)^{12} -
\left(\frac{\sigma}{r_{ij}}\right)^{6} \right],
\label{eq:hb}
\end{equation}
where ${\bf r}_{j}$ is the position vector of liquid particle $j$ with
momentum ${\bf p}_{j}$ and mass $m$, and $r_{ij}=|{\bf r}_{j} - {\bf
r}_{i}|$.

For simplicity we take the solute-solvent interaction to be a two-site
LJ potential in which each site of the molecule interacts with the
solvent atoms through precisely the same LJ potential:
\begin{equation}
\begin{split}
\phi({\bf R},{\bf r}) &= \sum_{j=1}^{N} 4\epsilon \left[
\left(\frac{\sigma}{r_{1j}}\right)^{12} -
\left(\frac{\sigma}{r_{1j}}\right)^{6} \right] + \\ & \sum_{j=1}^{N}
4\epsilon \left[ \left(\frac{\sigma}{r_{2j}}\right)^{12} -
\left(\frac{\sigma}{r_{2j}}\right)^{6} \right].
\end{split}
\label{eq:phi}
\end{equation}
Here $r_{ij}=|{\bf r}_{j} - {\bf R}_{i}|$.  In the application of the
mode-coupling theory we need to evaluate the derivative of $\phi({\bf
R},{\bf r})$ with respect to the reaction coordinate $R$.  This is
done using the chain rule, and the results is given by
\begin{equation}
\frac{\partial \phi({\bf R},{\bf r})}{\partial R} = \frac{1}{2}
\left\{\frac{\partial \phi({\bf R},{\bf r})}{\partial Z_{2}} -
\frac{\partial \phi({\bf R},{\bf r})}{\partial Z_{1}} \right\},
\label{eq:dphi}
\end{equation}
where $Z_{1}$ and $Z_{2}$ are the $z$ components of the position
vector of the diatomic molecule ${\bf R}_{1}$ and ${\bf R}_{2}$,
respectively, and the factor $\frac{1}{2}$ comes from the Jacobian.
The derivatives of $\phi({\bf R},{\bf r})$ along the other directions
vanish since the system is frozen along the $z$-axis.

To test our approach, we have calculated the static and dynamic input
required to obtain the time-dependent friction, using the Monte Carlo
and molecular dynamics simulation techniques.  While other approaches,
such as the integral equation theory and a proper mode-coupling
treatment of density fluctuations, can be used to obtain the static
and dynamic input, we feel that a fare test of our theory should rely
on numerically exact input.  Thus, despite the success of the
theoretical approach in predicting structural~\cite{Rabani01b} and
dynamical~\cite{Rabaniun} properties in LJ systems, we limit our study
to the more accurate simulation approach.

The static input was obtained using the molecular dynamics (MD) method
for $N=500$ particles (including the diatomic) and for a set of
densities $\rho=0.5,0.6,0.7,0.8,0.9,1.0$ in reduced LJ
units.\cite{AllenTildesley} Approximately $10^{6}$ MD steps were made
for each density.  Every $50$ steps we collected data for
$\zeta^{\dagger}(0)$, $\ddot{\zeta}^{\dagger}(0)$, $S(q)$ and for the
vertex $V^{\dagger}({\bf q})$.  The self-intermediate and intermediate
scattering functions were calculated using the molecular dynamics
method for identical conditions.  The results for each density were
averaged over $10$ different runs of total run time of $t=15$ reduced
LJ units.

\begin{figure}
\begin{center}
\includegraphics[width=8cm]{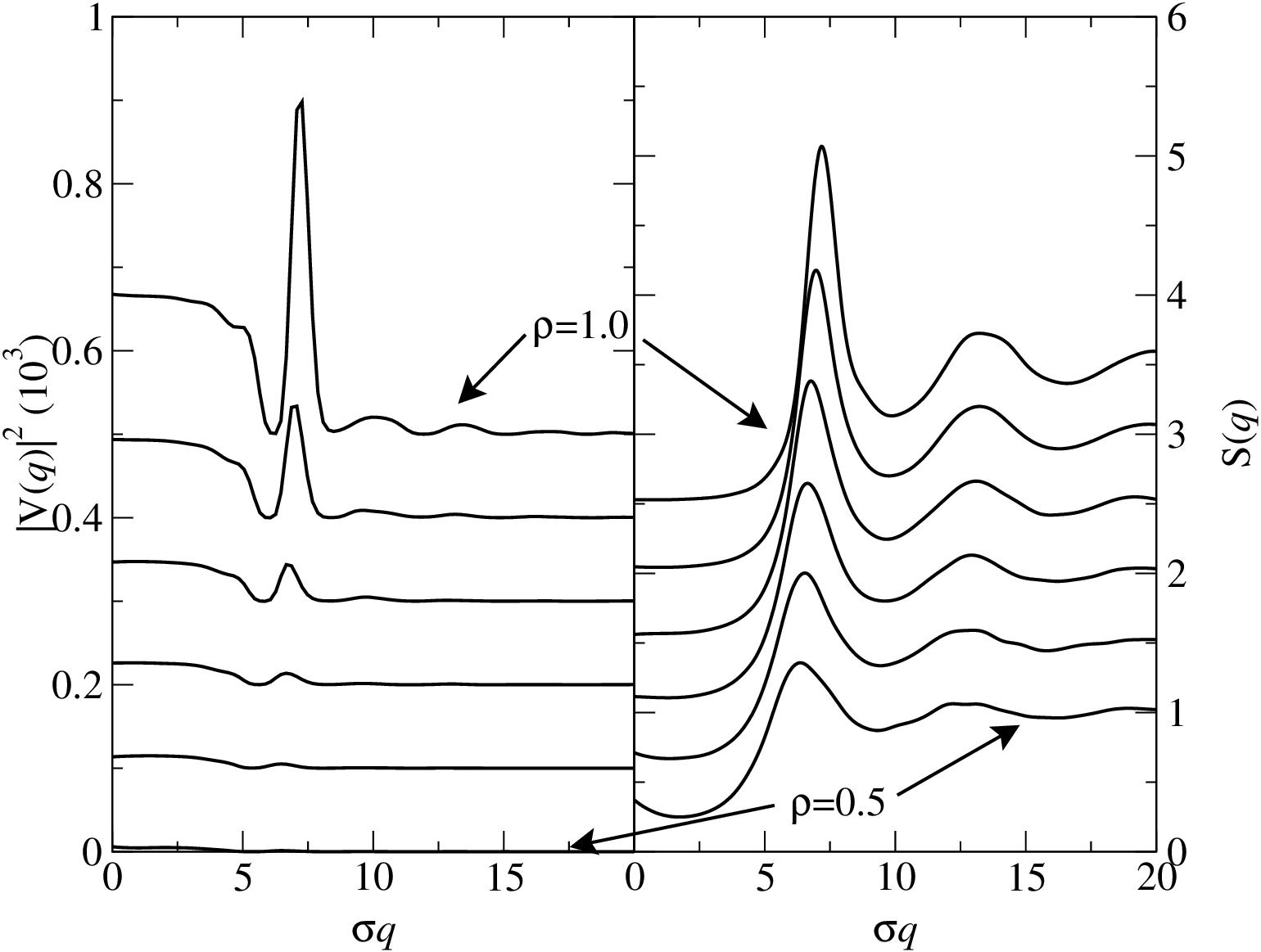}
\end{center}
\caption{The vertex square (left panel) and the static structure
factor (right panel) for different liquid densities
$\rho=0.5,0.6,0.7,0.8,0.9,1.0$ at $T=2.5$.  All results are in reduced
LJ units. For clarity, the results for different densities are shifted
vertically, from low to high densities.}
\label{fig:static}
\end{figure}

In Fig.~\ref{fig:static} we plot the static structure factor and the
vertex squared for all densities studied in this work.  The features
observed for the static structure factor are well understood, and have
been discussed elsewhere.\cite{HansenMcDonald} The magnitude of the
vertex given by $V^{\dagger}({\bf q})=\left\langle
\frac{\partial}{\partial {\bf R}^{\dagger}} [\phi({\bf
R}^{\dagger},{\bf r})-w({\bf R}^{\dagger})] b_{{\bf q},-{\bf q}}
\right\rangle$ determines the contribution of the different density
modes to the decay of the memory friction at intermediate and long
times.  We find that at high liquid densities the major contribution
comes from modes with a characteristic wavelength of $q \approx
2\pi/\sigma $.  This wavelength corresponds to the average
interparticle separation, and its value slightly increases with
decreasing density (note the shift in the position of the maximum of
$S(q)$ with density). Thus, density modes associated with liquid
motion on length scales of the interparticle separation (namely, on
length scale of $\sigma$) contribute the most to the friction at
intermediate and long times.

As expected, the contribution of lower wavelengths to the decay of the
memory friction becomes more significant at lower liquid densities.
In fact, at the lowest density studied, the contribution of density
modes below $q \approx 2\pi/\sigma $ to the decay of the memory
friction at intermediate and long times is more significant than the
contribution of modes near the first peak in $S(q)$.  We attribute
this effect to the change in the mechanism of self-diffusion of liquid
particles.  At high liquid densities, the self-diffusion is dominated
by opening of the cage surrounding a system, with a typical length
scale of $\sigma$ related to the size of the cage.  This is not true
at low liquid densities, where liquid particles can hop over much
larger distances.  In addition, we find that motion within the cage
become significant at higher liquid densities, as reflected in non
vanishing values of the vertex above $q \approx 2\pi/\sigma$.

\begin{figure}
\begin{center}
\includegraphics[width=8cm]{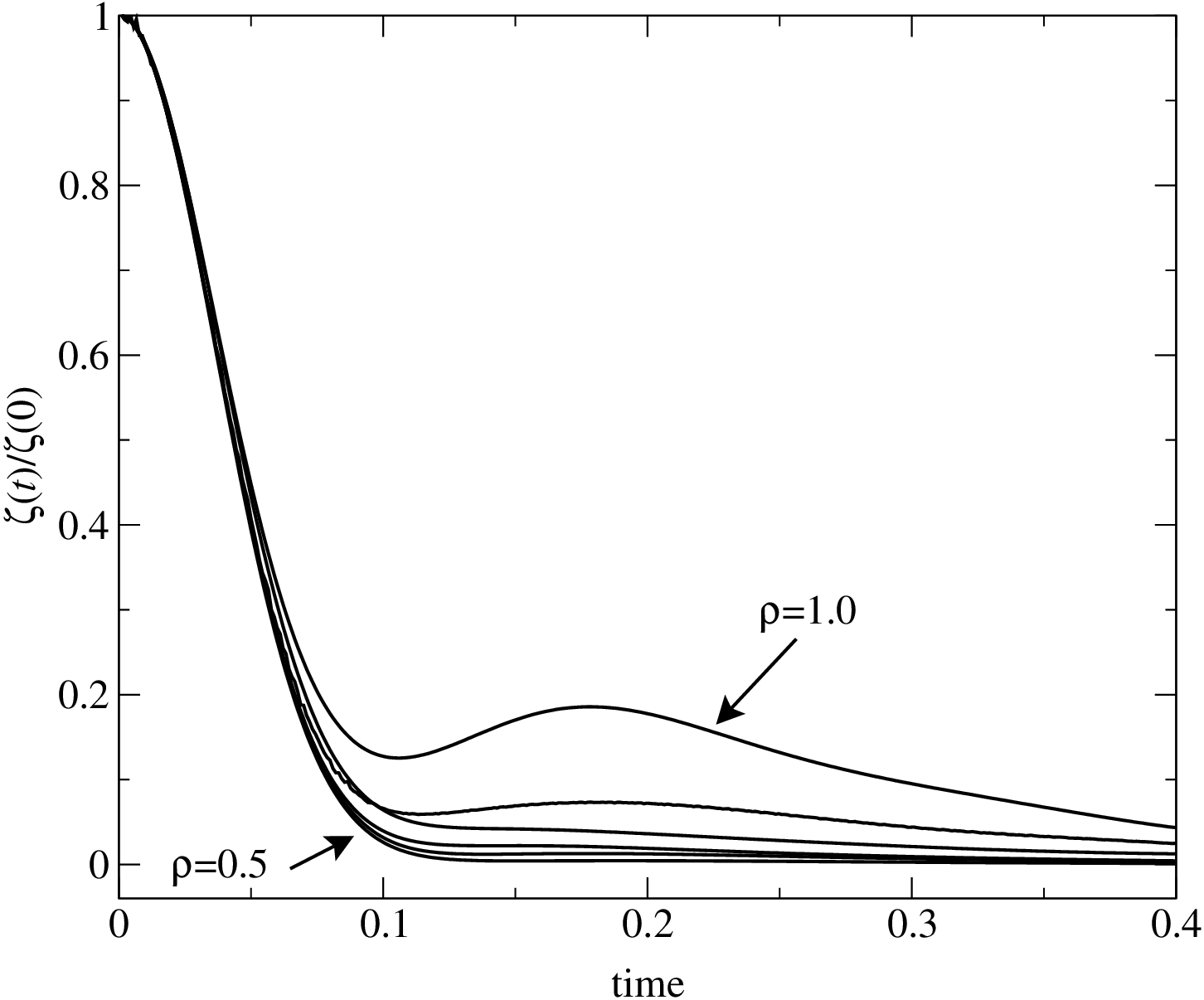}
\end{center}
\caption{A plot of the normalized memory friction $\zeta^{\dagger}(t)$
given by Eqs.~(\ref{eq:zetaB}) and (\ref{eq:zetamc}) for different
densities $\rho=0.5,0.6,0.7,0.8,0.9,1.0$ at $T=2.5$, in reduced LJ
units.}
\label{fig:zeta}
\end{figure}

Using the static input obtained from the Monte Carlo simulations along
with the values of the self-intermediate and intermediate scattering
functions obtained from molecular dynamics simulations, we have
calculated the total memory friction given by Eqs.~(\ref{eq:zetaB})
and (\ref{eq:zetamc}).  The results are shown in Fig.~(\ref{fig:zeta})
for all densities studied in this work.  As can be seen, the decay of
the memory friction is characterized by two time scales.  A fast decay
dominated by $\zeta^{\dagger}_{B}(t)$ followed by a slower decay
dominated by $\zeta^{\dagger}_{MC}(t)$.  The contribution of the
mode-coupling portion to the total memory friction is significant only
at high liquid densities.  At the lower densities studied the
mode-coupling portion of the memory friction is negligible.

For the specific model studied here where the solute-solvent
interactions equal to the solvent-solvent interactions, the memory
friction can also be obtained by inverting the GLE for the velocity
autocorrelation function of a neat fluid.\cite{Berne70a} Straub,
Borkovec, and Berne~\cite{Berne88} have calculated the velocity
autocorrelation function for the LJ fluid using the molecular dynamics
technique, and obtained the memory friction by inverting the proper
GLE.  Our results obtained from the mode-coupling theory agree well
with their simulated results (not shown).  Specifically, we find that
the decay of the memory friction at short times is nearly independent
on the liquid density, in agreement with the molecular dynamics
results.\cite{Berne88} Moreover, at high liquid densities, our theory
provides semi-quantitative agreement with the molecular dynamics
results at all times.  For low liquid densities, we observe small
deviations from the simulation results at intermediate times.  This
shortcoming of the mode-coupling approach is expected since the
simulated memory friction becomes slightly negative at intermediate
times, and the mode-coupling approximation is known to fail under such
circumstances. However, since the contribution of the mode-coupling
portion to the memory friction is relatively small at these low
densities, this has a vanishing effect on the value of the
transmission coefficient.

\begin{figure}
\begin{center}
\includegraphics[width=8cm]{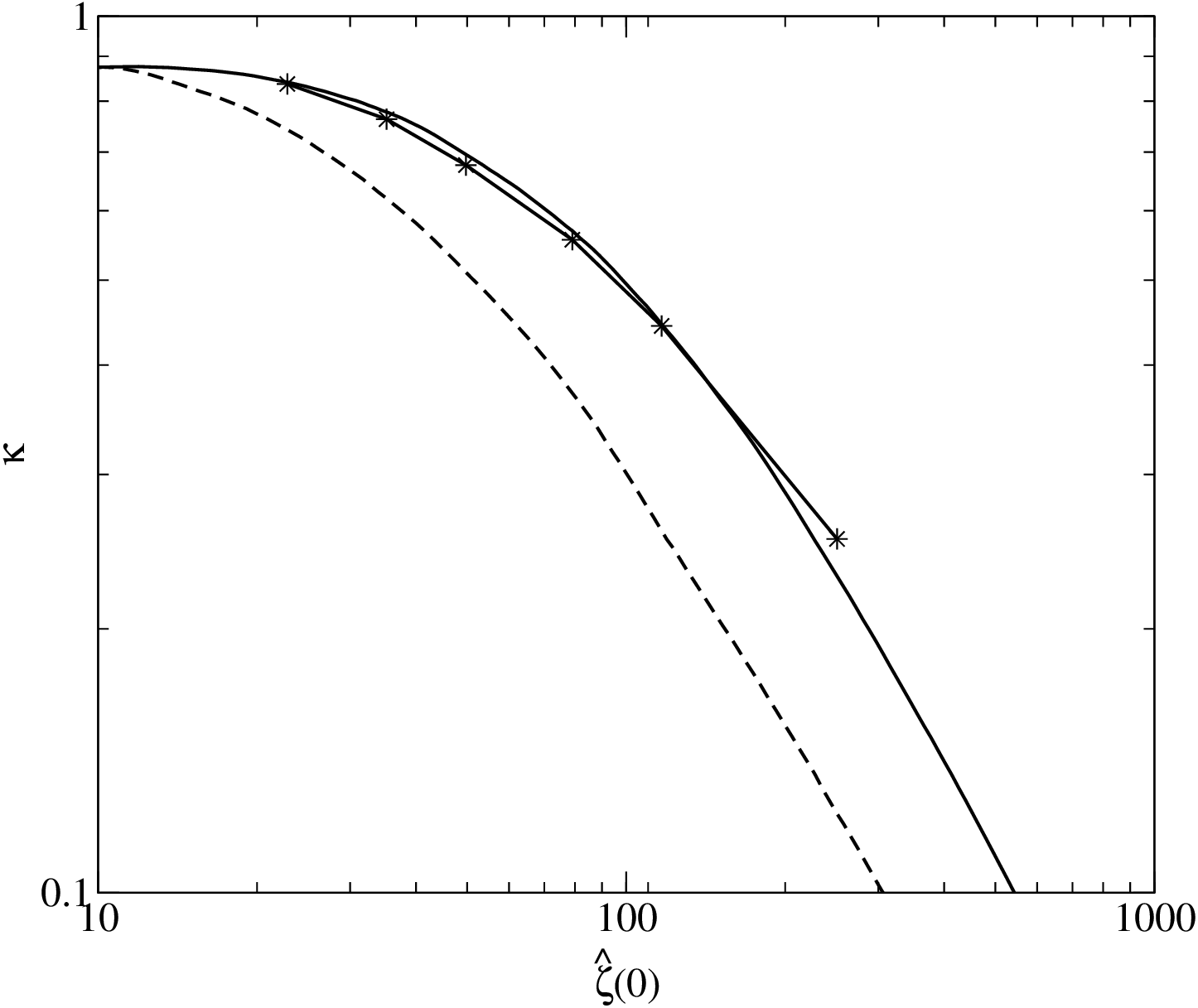}
\end{center}
\caption{The transmission coefficient as a function of the integrated
friction in the limit of spatial diffusion.  The stars connected by
solid line are the results of the present mode-coupling theory.  Solid
line is the result of the weak collision theory using an exponential
friction,\cite{Berne88} and the dashed line is the Kramer's result.}
\label{fig:kappa}
\end{figure}

Using the memory friction obtained from the kinetic and mode-coupling
theory we have calculated the transmission coefficient by solving
Eq.~(\ref{eq:ghrate}) self-consistently.  The results for $\kappa$ as
a function of the integrated friction $\hat{\zeta}^{\dagger}(0)$
(cf. Eq.~(\ref{eq:laplace})) are shown in Fig.~\ref{fig:kappa}.  We
compare our results to the results obtained by the weak collision
theory (based on a simple connection formula) using an exponential
friction,\cite{Berne88} and to Kramer's theory.\cite{Hanggi90} For the
present model the weak collision theory provides a quantitative
agreement for the transmission coefficient (within the noise level of
the simulations) in comparison with results obtained using the
absorbing boundary approximation~\cite{Berne85b,Berne85c} to the
reactive-flux formalism.\cite{Chandler78,Berne85} As can be seen, our
theory provides quantitative results for the transmission coefficient
over the entire range of frictions studied.

The different values of the integrated friction
$\hat{\zeta}^{\dagger}(0)$ at which we have calculated $\kappa$ where
obtained by scanning the liquid density from low $\rho=0.5$ to high
$\rho=1.0$ values.  For each density we have calculated the static and
dynamic input required by our mode-coupling theory, and obtained the
transmission coefficient by solving Eq.~(\ref{eq:ghrate})
self-consistently.  Alternatively, one can control the integrated
friction $\hat{\zeta}^{\dagger}(0)$ by changing the value of the
reduced mass $\mu$ of the isomerizing diatomic molecule
(cf. Eq.~\ref{eq:memorykernel}).  We find that our mode-coupling
theory provides similar quantitative results (not shown) for the
transmission coefficient when the reduced mass of the isomerizing
diatomic molecule is varied.

\section{Conclusions}
\label{sec:conclusions}
We have presented a theoretical approach for the calculation of
reaction rates in condensed phases based on the Grote-Hynes formalism.
A combination of kinetic and mode-coupling theories were developed to
obtain the memory friction required by the Grote-Hynes formalism.  The
approach was applied to study a model isomerization reaction of a
diatomic molecule in a LJ fluid.  Good agreement for the transmission
coefficient was obtained in comparison with the simulation results of
Straub, Borkovec, and Berne~\cite{Berne88} based on the reactive flux
formalism.

Unlike simulation techniques, our approach is a {\em theory} and thus
provides additional insight into the reaction dynamics in liquids.
For example, we showed that the contribution of the mode-coupling
portion to the decay of the memory kernel at intermediate times is
significant only at high liquid densities.  Thus, an accurate
description of the reaction dynamics at low liquid densities can be
obtained from a kinetic theory alone.  Furthermore, the mechanism for
the decay of the memory friction (which is reflected in the reaction
rate) is quite different at low versus high liquid densities.  At high
liquid densities the decay of $\zeta^{\dagger}(t)$ is dominated by
liquid modes with a length scale comparable to the separation between
the fluid particles, while at lower liquid densities motion on larger
length scales also contributes the decay of $\zeta^{\dagger}(t)$.
This is significant for the development of coarse grained models for
reaction dynamics.

We believe that our approach will be useful in other situations in
which simulation techniques are still limited.  For example, for
reaction dynamics in supercooled liquids that are characterized by
slow density fluctuations.  Or for liquid hosts that are characterized
by quantum mechanical susceptibilities.  Work along these directions
is currently underway.

\section{Acknowledgments}
\label{sec:acknowledgments}
This work was supported by The Israel Science Foundation founded by
the Israel Academy of Sciences and Humanities (grant number
31/02-1). The authors would like to thank David Reichman for
stimulating discussions and Oded Hod for support with the simulations.


\pagebreak
\end{document}